# Semimetals for high performance photodetection

Jing Liu[1], Fengnian Xia[2], Di Xiao[3], Javier García de Abajo[4,5], Dong Sun[6,7,*]

Affiliations:

[1]State Key Laboratory of Precision Measurement Technology and Instruments, School of Precision Instruments and Opto-electronics Engineering, Tianjin University, NO. 92 Weijin Road, Tianjin 300072, People's Republic of China

[2]Department of Electrical Engineering, Yale University, New Haven, Connecticut 06511, United States

[3]Department of Physics, Carnegie Mellon University, Pittsburgh, PA 15213, USA.

[4]ICFO-Institut de Ciencies Fotoniques, The Barcelona Institute of Science and Technology, 08860 Castelldefels (Barcelona), Spain

[5]ICREA-Institució Catalana de Recerca i Estudis Avançats, Passeig Lluís Companys 23, 08010 Barcelona, Spain

[6]International Center for Quantum Materials, School of Physics, Peking University, Beijing, China.

[7]Collaborative Innovation Center of Quantum Matter, Beijing, China.

*Email: sundong@pku.edu.cn

## *Abstract*

**Semimetals are being explored for their unique advantages in low-energy high-speed photodetection, although they suffer from serious drawbacks such as an intrinsically high dark current. In this perspective, we envision the exploitation of topological effects in the photoresponse of these materials as a promising route to circumvent these problems. We overview recent studies on photodetection based on graphene and other semimetals, and further discuss the exciting opportunities created by the topological effects, along with the additional requirements that they impose on photodetector designs**.

## *Current Status of Photodetection Technology*

Photodetectors (PDs) are pivotal optoelectronic components of modern communication and sensing systems that have become ubiquitous in our daily life. In the visible and near-infrared (NIR) wavelength ranges, PDs with high performance, low cost and high level of integration with electronics have already been developed and widely deployed, mainly leveraging elemental (e.g., silicon and germanium) and compound (e.g., gallium arsenide and indium phosphide) semiconductors[1,2]. Currently, the main technological bottleneck of photodetection lies in the low photon-energy range of the electromagnetic spectrum. Fast, highly sensitive PDs operated at room temperature are not widely available at low photon energies[2-7], despite the important role that they could play in a vast range applications, such as homeland security, night vision, gas sensing, motion detection, and bioassays, just to name a few[7,8]. Although some promising directions have been explored, they still suffer from significant drawbacks; for example, HgCdTe technology has been widely adopted, but it involves high manufacturing costs, is difficult to integrate and requires cryogenic operation[2,7]. Alternatively, bolometers can operate at room temperature, but their speed of operation is usually low (< 1 kHz)[9,10]. This situation can hardly satisfy the demands of novel applications, such as infrared spectrometry and imaging, thus urging the development of innovative approaches to achieve fast, broad-band, room-temperature, integrated photodetection in the mid-infrared and terahertz spectral ranges.

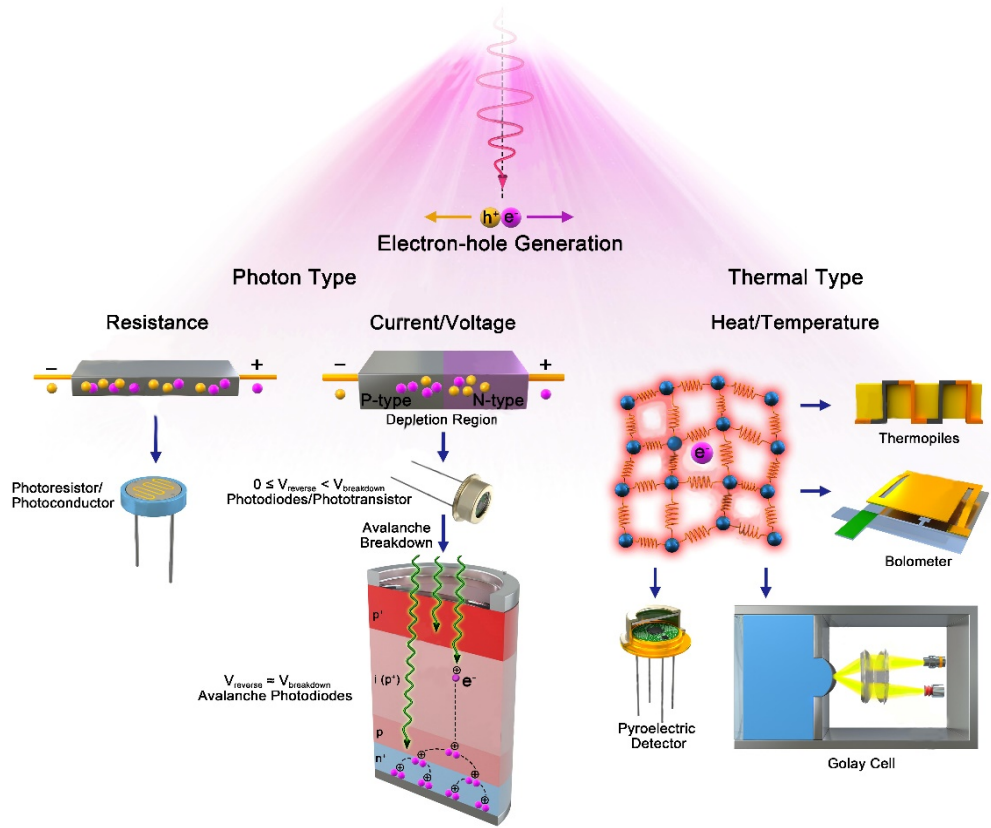

**Figure 1. Photodetection mechanisms.** Chart of currently used photodetection mechanisms and representative photodetector devices. Probed light produces hot charge carriers that directly change the resistance, generate a current, or produce a voltage in photon-type detectors; while they thermalize to increase the temperature of the material in thermal-type detectors.

Photodetectors leverage different light-induced effects that are readable through electrical measurements[1]. Figure 1 summarizes the most widely used photodetection mechanisms, classified according to the nature of the physical effect caused by the incident radiation: *photon-type* detectors rely on the direct production of photoexcited electron-hole pairs, while *thermal-type* detectors rely on the change in electron and/or lattice temperature upon thermalization of those hot carriers. A comparison of the performance of different representative thermal- and photon-type PDs that are suitable for low-photon-energy detection is also presented in Table 1. In a thermal-type detector, the absorbed incident radiation changes the material temperature, which in turns results in measurable changes of physical quantities that are probed electrically. Specifically, one can measure the temperature-induced variation of resistance in a bolometer[2,7], the modification of the voltage in pyroelectric materials[11] and thermopiles[12], or even gas expansion in Golay cells[13]. In general, thermal-type detectors are wavelength insensitive, slow and usually less expensive than their photon-type counterparts. In contrast, in the latter the electrical output signal comes directly from the charged photocarriers produced upon the photoexcitation, leading to improved signal-to-noise ratio and fast response, although they are more expensive and usually require cryogenic cooling for low-energy photon detection in order to minimize thermally generated charge carriers that compete with the targeted optically excited carriers by generating excessive noise otherwise.

**Table 1.** Performance comparison of representative *photon-* and *thermal-type* mid-infrared photodetectors.

| Figures of Merit | Photon Type (PN/PIN) | | | Thermal Type (Bolometers) | | |
|---|---|---|---|---|---|---|
| | InSb | HgCdTe | Type II InAs/GaSb | VO$_x$ | α-Si | Superconductor (HEB) |
| Detectivity (Jones) | ~$10^{11}$ @ 77 K | $10^{12}$ @ 77 K | $10^{11}$ @ 77 K | $10^7 - 10^8$ @ 300 K | $10^6 - 10^8$ @ 300 K | $10^{11} - 10^{12}$ @ 4.2 K |
| Response Time | ~ 30 ns | ~ 500 ns | ~ 100 ns | ~ 3 ms | ~ 300 ms | ~ 50 ps |
| Spectrum | 3 – 5 μm | 1 – 14 μm | 3 – 5.4 μm | 7-14 μm | 7-14 μm | 150-3000 μm |
| References | 1, 2 | 1, 2 | 1, 2 | 14 | 15 | 16 |

In this perspective, we focus our discussion on photon-type detectors because of their overall better performance relative to thermal-type. The most widely used photon-type detection scheme relies on the well-konwn p-doped-semiconductor/insulator/n-doped-semiconductor (PIN) structure illustrated in Fig. 2a. A reverse-biased PIN diode has a negligibly small dark current. When a photon of sufficient energy is absorbed in the depletion region of the diode, it creates an electron-hole pair. Then, the reverse-biased field sweeps these carriers away from that region, thus creating a photocurrent. Commercially available PIN photodiodes can reach quantum efficiencies of ~80-90% at the telecom wavelength (~1550 nm), while they are remarkably compact and highly integrable and operate at high speed. In the mid-infrared, matured photon-type photodetection technologies are based on HgCdTe and InAs, with specific detectivity ($D^* = \sqrt{A \Delta f}/\text{NEP}$, defined in terms of the noise-equivalent power (NEP), the detection area $A$ and the frequency band width $\Delta f$) above $10^{11}$ Jones (= cm$\sqrt{\text{Hz}}$/W) and operation response times in the nanoseconds range. Nevertheless, as mentioned above, these detectors usually require cryogenic cooling, which makes them bulky, expensive, power hungry and delicate[7]. Therefore, high performance, low-energy photon detectors that can perform well without cryogenic cooling remain a highly desirable challenge.

To achieve a device performance that circumvents the existing technological bottleneck, a strategy consists in searching for new material platforms, on which strong efforts are being invested within the materials science and optoelectronic communities[8,17-25]. In this line, this perspective portraits exciting opportunities on photodetection enabled by semimetallic materials, which are promising candidates to achieve highly sensitive, low-energy photodetection at ultrafast operation speed.

## *Semimetals vs Semiconductors for Photodetection*

Semimetallic materials have not been traditionally considered as candidates for photodetection because of the obvious drawback of the high dark current associated with the current flow through them when applying a bias voltage, which reduces their detection sensitivity. However, the demonstration of a graphene-based field-effect transistor (FET) photodetector a decade ago[26] revealed the feasibility of using metallic materials in high speed photodetection applications over a broad wavelength range, despite their relatively high dark current. This work triggered much research on semimetal-based photodetection. Although the responsivities of semimetal-based detectors are usually far from optimal, their gapless electronic structures endow them with unprecedented broadband photoresponse down to the far infrared[27-30], combined with extremely fast operation speed [27,31-33], both of which directly address the mentioned technological bottlenecks. As illustrated in Fig. 2b, d, the limitation on detectable photon energy imposed by an energy gap is absent in semimetals, and thus, the detectable photon energy range extends naturally down to the low energy end. Additionally, the transient lifetimes of the photoexcited carriers, which are relatively long in semiconductors and therefore compromise the operation speed of semiconductor-based PDs, are dramatically reduced in the absence of a bandgap through rapid electron-electron scattering[34,35], thus boosting the operation speed of semimetals.

Despite these salient advantages, semimetal-based PDs require unbiased operation in order to reduce the dark current, and therefore, in contrast to semiconductor detectors, charge separation cannot be achieved through the reverse bias approach discussed above for PIN diodes. Consequently, semimetal PDs operating in an unbiased photovoltaic mode (Fig. 2c) need to achieve charge separation through other less efficient mechanisms, such as a built-in electric field, the photothermoelectric effect or the photo-Dember effect[24,26,32,36-38]. The short transient lifetime of photoexcited carriers in semimetals further aggravates the charge separation problem (i.e., the time window for charge separation is reduced), rendering it rather inefficient and producing a low photodetection responsivity. The need for fast charge separation mechanisms in semimetals is thus critical.

Two additional factors may further affect the performance of semimetal PDs. The dark noise in a typical metal-semimetal-metal FET device is dominated by thermal agitation of the charge carriers[39,40], even if its dark current is small in the absence of an external bias. Also, the photocurrent of an unbiased photodetector usually exhibits a turn-on threshold (i.e., the light intensity needs to exceed a certain threshold power to produce a detectable photocurrent in an unbiased device, see Fig. 3c), which can be significant without external bias, thus imposing a severe constraint on the lowest detectable light intensity of a semimetal-based PD.

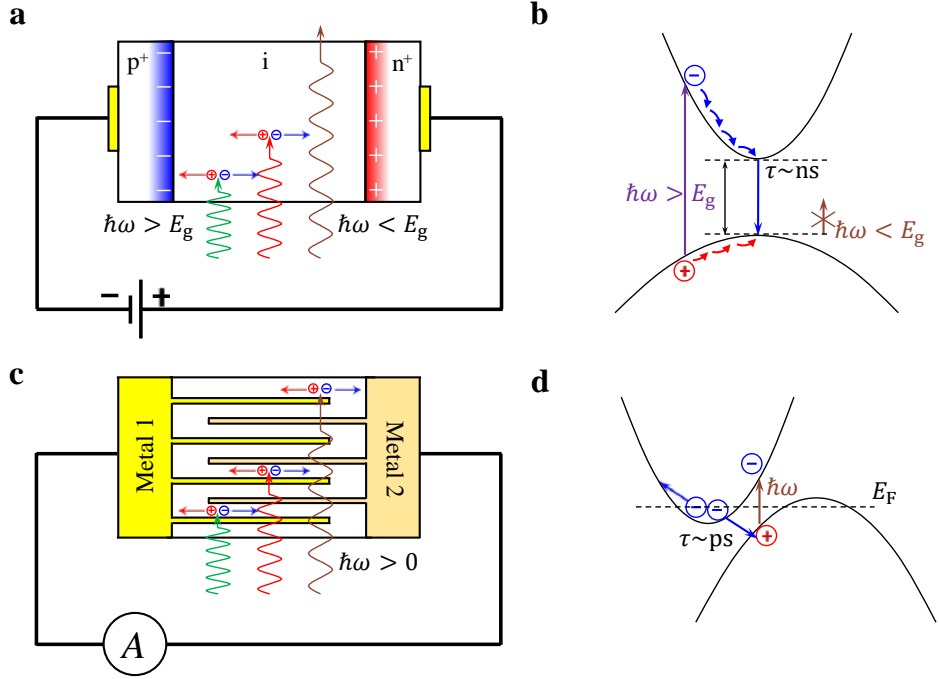

**Figure 2. Semiconductor- vs semimetal-based photodetectors. a,** Sketch of PIN photodiode under reversed bias. Photons of energy $\hbar\omega$ exceeding the band gap $E_g$ can be absorbed and detected. **b,** Illustration of the generation and recombination of photocarriers in semiconductors. Hot carriers generated with $\hbar\omega > E_g$ exhibit a typical relaxation time of the order of nanoseconds through electron-hole recombination across the bandgap. **c,** Sketch of a typical cross-finger-type field-effect transistor for photodetection based on semimetals. Light can be absorbed and detected for arbitrarily low energy, without the limitation imposed by a band gap. **d,** Illustration of the generation and recombination of photoexcited carriers in semimetals. Fast recombination through electron-electron scattering occurs within picoseconds.

Because of the above, a suitable figure of merit that quantifies the sensitivity of a photodetector has to be defined, other than the NEP, because the turn-on threshold power usually dominates the NEP in semimetal-based PDs and can by far exceed the equivalent power converted from both dark noise and the device photoresponsivity. However, if we come back to the definition that NEP is the required optical power that gives a signal-to-noise ratio of one in a one hertz output bandwidth, other related figures of merit derived from NEP, such as the specific detectivity D*, stay the same.

## *Development of Semimetal-Based Photodetectors*

Besides the advantages of fast and broadband operation, early developments of semimetal-based PDs benefited from several unique properties of graphene: the linear electron dispersion and atomically thin layered structure contribute to carrier multiplication that increases the photodetection quantum efficiency[41-44] and leads to large photothermoelectric response[32,37,38] with high mobility[45,46]; additionally, graphene is CMOS compatible[47], integratable with flexible electronics and amenable to hybridization with other materials. Over the last decade, these properties have been capitalized for photodetection, as summarized in recent reviews[8,24,48-51].

Despite these successful demonstrations, the semimetallic band structure and unbiased operation of this material limit the photoresponsivity of graphene PDs. Solutions to this problem have been sought in using photonic and plasmonic cavities to boost the near-field enhancement, as well as hybridization in van der Waals material heterostructures[24]. These approaches can improve the responsivity, but just at the expense of reducing the broadband response and ultrafast operation. For example, boosting the responsivity through field enhancement in optical cavities generally relies on narrowband resonances that jeopardize broadband operation[52-58]. Interestingly, by hybridizing with other materials, the latter not only act as light harvesters, but also provide interfaces and hetero-junctions that can facilitate the separation of photoexcited electron-hole pairs. A wide range of materials has been explored in this direction, including polymers[59], zero-dimensional (0D) quantum dots (QDs)[60-62], one-dimensional (1D) nanorods and nanotubes[63,64], two-dimensional (2D) layered materials[65-68] and bulk (3D) semiconductors[28,52,53,69-72], reaching measured responsivities as large as 435mA/W[71], but compromising speed and spectral range (i.e., the main advantages of semimetals).

Beyond the responsivity, the sensitivity of the detector also depends on noise current and quantum efficiency (QE). Unfortunately, the noise current is underestimated in a significant fraction of published literature by calculating the shot noise limit, instead of experimentally measuring it, as extensively discussed in a commentary[73]. Compared to the responsivity, the QE provides a better characterization of how efficiently photons can be converted into an electrical current. A detector with low QE can reach very high responsivity if it responds slowly enough to integrate the generated current over a long period of time. The reported high internal QE reaching 30%–60% in graphene PDs[74,75] presumably originates in carrier multiplication, which may enhance the QE by a significant factor[76,77]. However, carrier multiplication is more efficient at larger photon energy, so it does not improve low-energy photon detection significantly. This large internal QE further indicates that the observed low response is not limited by the intrinsic performance of graphene, but rather by other factors such as non-optimized device layouts and electrical contacts.

Recent advances in improving the responsivity are enabled by the rapid development of topological semimetals[78,79]: materials with semimetallic character possessing topologically nontrivial electronic band structures. The experimental realization of both Weyl and Dirac semimetals[80-84] has brought the field to the forefront of quantum condensed-matter research, although early studies date back to the quantum-Hall effect in the 1980s[85,86], while topological insulators well extensively studied in the 2000s[87-89]. A straightforward approach consists in replacing graphene (a two-dimensional Dirac semimetal) by three-dimensional Dirac semimetals, such as $Cd_3As_2$, which can improve the responsivity by about one order of magnitude without losing broadband response and ultrafast operation speed[27], although in this case such improvement actually bears no connection with the topologically nontrivial band structure of $Cd_3As_2$ and is simply produced by the enhanced absorption of the bulk material compared to the 2.3% value of monolayer graphene. Additionally, using the shift current response of Weyl semimetals, a boost of at least two orders of magnitude in responsivity is expected by benefitting from the Berry field enhancement at the vicinity of Weyl nodes, which is a purely topological effect[90]. We discuss this issue in more detail below and note that such enhancement should not affect the ultrafast operation speed and broadband response, although further experimental verifications are still needed.

## *Topology and Photodetection*

Certain topological properties of topological semimetals can greatly improve the photoresponse of PDs. Take Weyl semimetals as an example. These materials host Weyl fermions traveling parallel or anti-parallel to its spin moment, which defines the chirality of a specific Weyl cone. The energy of a Weyl fermion is proportional to its momentum, forming a cone-like structure in energy-momentum space (Fig. 3a). Most importantly, each chiral Weyl node can be viewed as a 'monopole' of the Berry flux field, an effective magnetic field in momentum space (Fig. 3b)[91]. These magnetic monopoles have direct effects on the electron motion and result in various intriguing topological effects. One of such effects is on the shift current response[90,92-99], which results from the shift of charge center during interband photoexcitation in non-centrosymmetric materials excited by linearly polarized light[100], and constitutes an intrinsically different way of generating a photocurrent compared with semiconductor p–n junctions, where a built-in electric field separates electrons and holes. The shift of the charge center can be expressed as a change of the Berry connection[90], a vector potential that generates the Berry flux field and the phase of the velocity operator, and consequently, the corresponding conductivity tensor is expected to be greatly enhanced when the excitation takes place in the vicinity of Weyl nodes, where the Berry flux field diverges (Fig. 3c). This topological enhancement effect has been recently been experimentally verified in both Type-I TaAs and Type-II $TaIrTe_4$ [101,102].

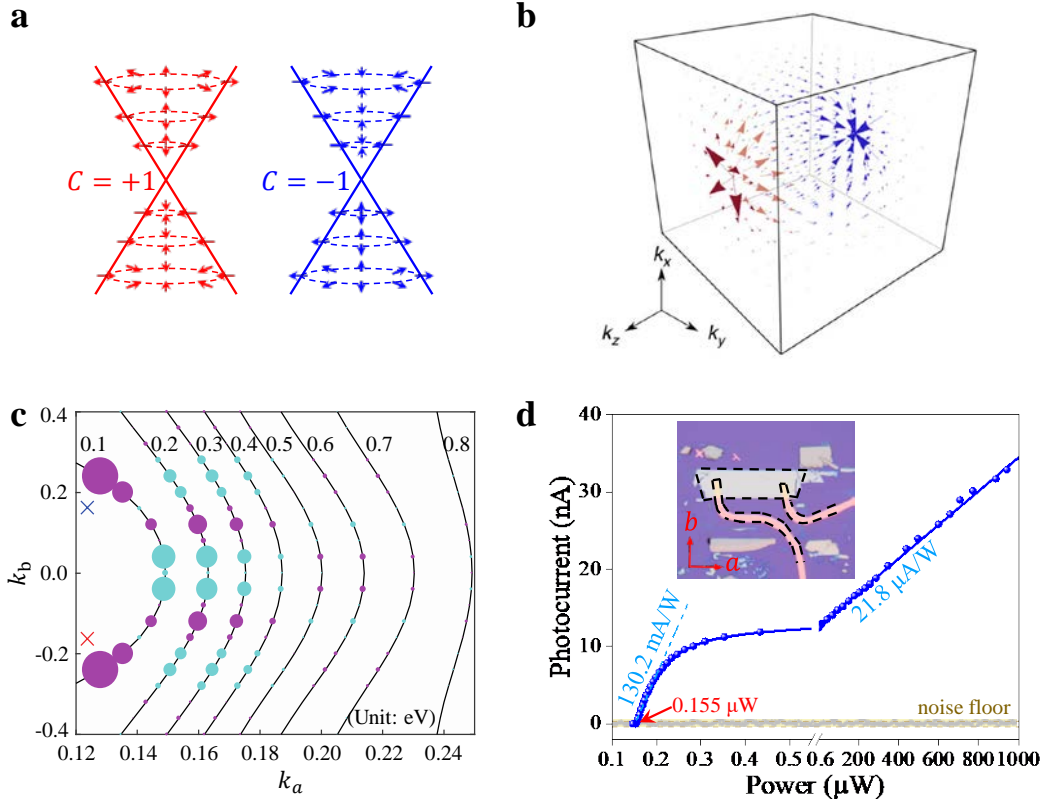

**Figure 3. Topological enhancement of the shift current response in Weyl semimetals**. **a,** Sketch of a pair of Weyl cones along with their spin texture indicated by the arrows. Red and blue lines and arrows denote opposite chirality. **b,** Vector plot of the Berry curvature of Weyl semimetals in momentum space. A pair of Weyl nodes emerges, hosting each a monopole, here illustrated through arrows showing the flux of the Berry curvature between them. **c,** Integrand of effective third-order optical conductivity tensor shown along contours of fixed electron energy in the material. The disc radius indicates the absolute value of the matrix elements; purple (cyan) color denotes positive (negative) sign. The red (+1) and blue (-1) crosses indicate the Weyl Points with opposite chirality. **d,** Light-power-dependence of the photocurrent at 4 μm incident wavelength and linear polarization along a-axis excitation. The device is marked by dashed lines in the optical microscope image shown in the inset. Figures reproduced with permission from: b, ref 93, 2016 Nature Publishing Group; c,d ref. 101, 2019 Nature Publishing Group.

Figure 3d shows the power dependence of the photoresponse toward 4-μm light in a typical $TaIrTe_4$-based FET device[101], which strongly differs from that of semiconductor PIN detectors. First, the $TaIrTe_4$ detector shows a clear turn-on threshold at low excitation power of ~155 nW (we note that the turn-on threshold is usually overlooked in graphene PDs due to their much smaller responsivities). Second, above the turn-on threshold the photocurrent experiences a dramatic change with excitation power, arising from the topologically enhanced photoresponse, quantified by a linear responsivity of 130.2 mA/W. This giant response saturates at very low power, indicating a narrow linear dynamic range (LDR), but this should not affect its application in highly sensitive photodetection. Third, after the transitional power range, the response saturates and reaches a second linear dynamical range with a relatively low responsivity of 21.8 µA/W. In this power region, the response is dominated by topologically trivial effects such as photothermoelectric.

The large responsivity obtained at low powers indicates that charge separation due to the shift current response is very efficient. An additional advantage of this mechanism is it's an intrinsically ultrafast behavior, thus facilitating to overcome the low responsivity problem of a semimetal-based PDs. However, the turn-on threshold, resulting from potential barriers in the device, as demonstrated on a prototype device illustrated in Fig. 3, can become an issue that limits detectivity. Such barriers are primarily determined by the work function difference at metal-semimetal interfaces in a FET device, but they can also relate to other factors such as impedance match with external circuits. The potential barrier can be lowered down through device engineering by carefully matching the work function of a metal contact and the doping level of a semimetal, although constructing PN junctions is perhaps a more feasible approach, because the doping levels of the P/N regions can be modified separately to achieve more flexible control on the junction. We also note that even with the fixed potential barriers, the turn-on threshold power can depend on excitation wavelengths because higher-energy photons excite carriers with larger kinetic energy, for which barrier reflection is reduced, and consequently, so is the turn-on threshold power.

## *Technical Challenges*

Despite the excitement created around topological enhancement leading to a large photoresponsivity in Weyl semimetals, several technical challenges need to be addressed, as we discuss in this section.

### *Determination of the final "silicon" type of hero material*

Successful functional devices generally rely on high-quality material growth, but topological semimetals are still difficult to produce, and their wafer-scale CMOS compatibility is also a challenge. Recently, theorists have exhaustedly explored topological candidates from the >26,000 available non-magnetic crystalline compounds and compiled the results in the International Crystal Structure Database[103-105] (ICSD), where thousands of new topological materials were proposed. Additionally, new topological candidates can arise from artificial structure engineering of existing materials, such as stacking of twisted layered materials[106-108]. It is obviously unfeasible to exhaustively investigate such large number of material candidates experimentally, and thus, in order to narrow down the scope of promising candidates, we provide some key considerations to take the advantages of the abovementioned Berry phase enhancement effects on the shift current response. These considerations are beyond the common requirements on wafer scale epitaxial growth and general material properties, such as air-stability, non-toxicity, CMOS compatibility and easy on-chip integration. First of all, topological enhancement of the shift current response requires materials with inversion symmetry breaking, such as Weyl semimetals and chiral-fermion materials[109-114]. Although a Dirac node is also a singularity point of the Berry curvature, similar to a Weyl node, inversion symmetry forbids shift current generation in Dirac semimetals. Consequently, Dirac semimetals do not share similar topological enhancement. In contrast, chiral fermion materials are likely more favorable for shift current generation compared to Weyl semimetals because Weyl cones with opposite chirality are no longer energy degenerate, so a wider photon energy region undergoing possesses topological enhancement may be covered. Second, the Fermi level has to be close enough to the nodal points to allow the topological effect to come into play with related optical transitions. If the nodal points are also far away from the topologically trivial bands, it further helps to determine the response unambiguously from the topological nontrivial bands. In addition, the cones should be well separated in the momentum space, so that it is possible to address them separately through optical transitions. Third, suitable dopants have to be identified in order to suitably tailor material doping. Doping and its tunability are not only critical to optimize the potential barriers with metal electrodes (i.e., to reduce the turn-on threshold power), but they also help to tune the wavelength range over which topological enhancement takes place[115]. Fourth, layered materials can be conveniently integrated with other 2D layered materials for actual applications[116], for example to prevent degradation by BN capping if the material environmentally unstable. Incidentally, BN provides an ideal dielectric layer that can play the role of $SiO_2$ in silicon technology[117].

*Device designer's perspective*

The absence of an external voltage bias and the involvement of topological effects demand special considerations on the device designs. In particular, the absence of an external bias implies that one has to minimize the turn-on threshold power of a FET photodetector in order to increase its sensitivity, which in turn requires an engineering effort to lower the semimetal-contact barrier by conventional approaches such as semimetal doping or choosing a metal contact with a suitable work function[118]. Alternatively, one could combine optical approaches to solve this problem (e.g., optically bias the device just below the turn-on threshold). Furthermore, solutions may also come from a new topological effect, as we discuss below.

A second issue is related to the unbiased device structure: without an external bias, other efficient charge separation mechanisms need to be found. Although the shift current response provides an efficient mechanism for charge separation, it imposes additional symmetry requirements, which require careful device engineering. In this respect, the photoresponse observed in Type-I TaAs is a second-order nonlinear optical effect[102], while the photoresponse observed in Type-II TaIrTe$_4$ is a third-order effect, equivalent to a shift current response under an in-plane DC electric field[101]. For a second-order nonlinearities (e.g., in TaAs), the correct crystal facet with suitable symmetry has to be selected to provide a non-vanishing shift current response. For third-order nonlinearities (e.g., in TaIrTe$_4$), a DC electric field is needed to provide a nonzero shift current response under normal incidence along the crystallographic c-axis, as otherwise the shift current response cancels due to $C_2$ symmetry in a $C_{2v}$ crystal. This limits the photocurrent response to a region in which a DC electric field exists after photoexcitation, such as along the semimetal-electrode interface. As a result, special device structures (e.g., a cross-finger device geometry, see Fig. 2c) need to be fabricated[118].

The third issue relates to engineering the photon energy range, and this can benefit from topological enhancement. In this respect, because the Berry field diverges at the Weyl nodes, the closer an optical transition is to a Weyl node, the more Berry field enhancement takes place in the shift current response. However, this argument is based on the assumption that the Fermi level goes through the Weyl nodes, while theory informs us that this only applies to type-II Weyl semimetals[102]. Additionally, if the Fermi-Dirac carrier distribution is modified by either temperature or impurity doping, the divergence of the shift current response at a nodal point can be smeared out or truncated, and the energy profile of topological enhancement be modified[102]. This implies that one can tune the range of topological enhancement by actively controlling the doping level. Incidentally, in this approach, the temperature needs to be stabilized to preserve a constant response.

*On-chip topological integration*

Due to wide range of existing layered topological semimetal spices, topological semimetals can be conveniently integrated in waveguides or cavities to boost light-matter interactions. Alternatively, layered species could be artificially stacked to generate exotic hybrid materials, so that different functionalities can be added, instead of simply improving the optical and photodetection performance. Novel opportunities for topological materials might come from the rich phenomenology of surface states and interface effect, which make it possible to carefully engineer edges and interfaces to achieve multiple functionalities[87,119]. Integration is no longer simply adding existing functions of multiple devices together, but, more importantly, generating new previously non-existing functionalities. We anticipate that topological materials will define their own integratable platforms, resulting in much improved optoelectronics devices.

## *Opportunities in Topologically Enhanced Photodetection*

Future solutions to the technical challenges of low-energy photodetection main benefit from the opportunities opened by topological physics in semimetals. For example, to address the important turn-on threshold problem in semimetal detectors, one could look for topologically protected dissipationless conducting channels to drive photoexcited carriers from the semimetal and the electrodes, instead of lowering down the contact-semimetal potential barrier though work-function matching[87,89,120,121]. In addition, the shift current response, which does not require an electric field built from potential difference, provides a well-fitted charge separation mechanism in such design [90,101,102].

Symmetries in topological materials could also be exploited by carefully engineering their edges and interfaces. For example, with specific symmetry breakings on an edge fracturing along a certain crystallographic direction, a type-II Weyl semimetal with $C_{2v}$ symmetry such as $WTe_2$ can efficiently realize charge separation to obtain a significant photocurrent response along the edge through the non-local Schottky Rohm mechanism[122]. Another possibility, yet to be verified, consists in using an exotic edge state (e.g., the Fermi arc of a Weyl semimetal) to further enhance light-matter interaction on an edge. In particular, topological edge states not only add extra conducting channels on top of the bulk conductance, but also penetrate into the bulk, thus enlarging the charge separation region and the magnitude of the current response.

Additionally, topological effects may provide novel control mechanisms on specific quantum degrees of freedom, such as the chirality-related circular selection rule of Weyl cones, which can help to distinguish the helicity of the light to produce helicity-sensitive PDs based on Weyl semimetals and chiral-fermions materials[93,123,124].

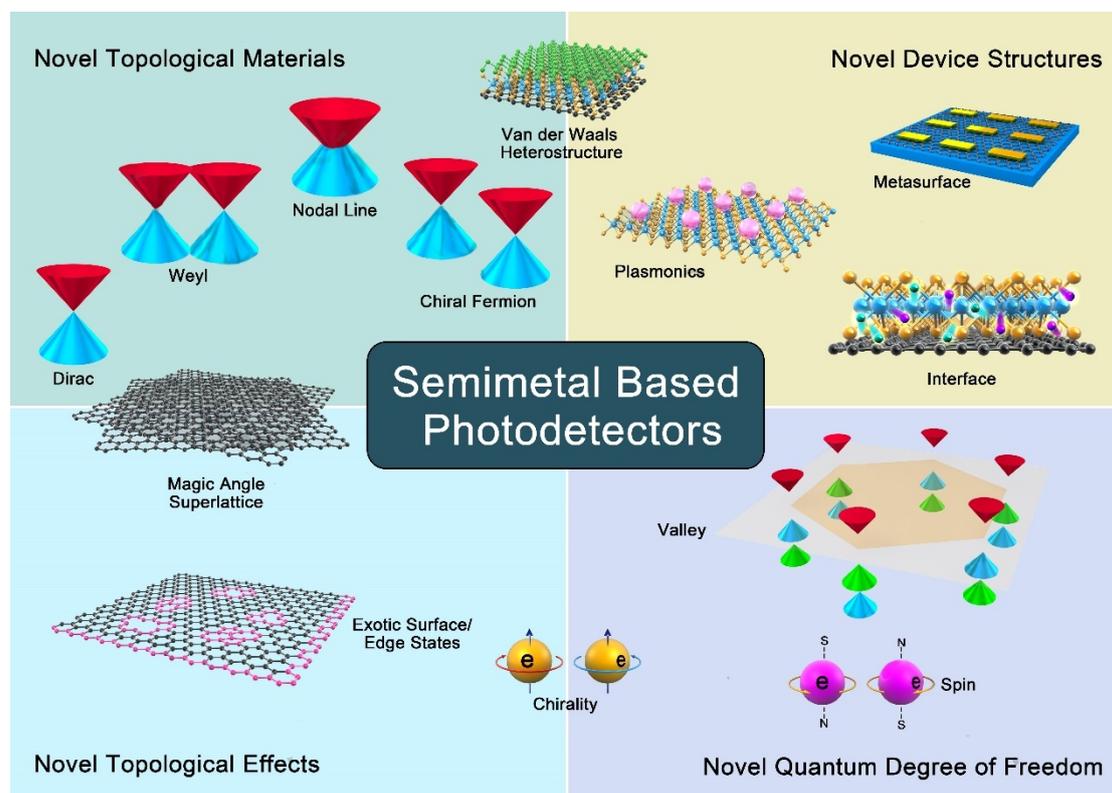

**Figure 4. Opportunities in semimetal-based photodetection.** Four different directions are here identified by exploiting novel topological materials, device structures, topological effects, and quantum degrees of freedom (left to right and top to bottom).

As a blooming field, topological effects of semimetals provide an exciting platform for new developments in photodetection. The interplay of chirality, quantum geometric effect, and exotic surface states with unconventional quantum degrees of freedom warrant plenty of interesting optical effects to be discovered with potential for unprecedented applications. Stacking of van der Waals materials offer an additional degree of freedom to customize semimetal materials to the required needs. We illustrate several of these possibilities in Fig. 4. With the existing interest on the enhanced shift current response of Weyl semimetals by topology, we foresee topological solutions to current technical issues in photodetection, which may well become the first commercial application of topological physics.

**Acknowledgement**

D.S. acknowledges support from the National Natural Science Foundation of China (NSFC

Grant Nos. 11674013, 91750109); D.X. acknowledges support from the Department of Energy, Basic Energy Sciences, Grant No. DE-SC0012509; J.G.A. acknowledges support from the Spanish MINECO (MAT2017-88492-R, SEV2015-0522) and ERC (Advanced Grant No. 789104-eNANO). J.L. would like to acknowledge the financial support by National Key R&D Program (2018YFA0307200) and the 111 Project (No. B07014).